\documentclass[11pt]{article}
\usepackage{epsfig}
\usepackage{amsmath}
\usepackage{amssymb}

\def\bea{\begin{eqnarray}}
\def\eea{\end{eqnarray}}
\def\beq{\begin{equation}}
\def\eeq{\end{equation}}

\def\be{\begin{equation}}
\def\ee{\end{equation}}

\def\5{\overline 5}

\begin{document}

\begin{center}

\thispagestyle{empty}

\begin{center}

\vspace{1.7cm}

{\LARGE\bf Enhanced gravitational scattering from large extra dimensions}

\end{center}

\vspace{1cm}

{Kazuya Koyama, Federico Piazza and David Wands}

\vspace{.8cm}
{\em }

{\em Institute of Cosmology and Gravitation, University of Portsmouth,} \\
{\em Mercantile House, Portsmouth PO1 2EG, United Kingdom}\\


\end{center}

\vspace{3cm}

\centerline{\bf Abstract}
\begin{quote}
 
We investigate whether enhanced gravitational scattering on small
scales ($<$ 0.1mm), which becomes possible in models with large extra
dimensions, can establish statistical equilibrium between different
particle species in the early Universe. We calculate the classical
relativistic energy transfer rate for two species with a large ratio
between their masses for a general elastic scattering cross section.
Although the classical calculation suggests that ultra-light WIMPs
(e.g., axions) can be thermalized by gravitational scattering,
such interactions are considerably less efficient once quantum
effects are taken into account on scales below the Compton
wavelength.
However the energy transfer rate in models with several
extra dimensions may still be sensitive to trans-Planckian physics.
\end{quote}

\vfill 

\newpage
\section{Introduction}

Higher dimensional models suggest a phenomenologically interesting
solution to the hierarchy problem.  A higher dimensional Planck mass
$M_D$ just above the electroweak scale is in fact compatible with the
observed weakness of four-dimensional gravity ($M_P\sim10^{19}$GeV) if
we consider relatively large compactification radii. In braneworld
models where gravity freely propagates in $4+d$ dimensions \cite{ADD}
and the $d$ internal dimensions share the same compactification radius
$R$, the latter is related to $M_D$ as
\begin{equation}
R \ \simeq \ \left(\frac{M_P}{M_D}\right)^{2/d}\, M_D^{-1} \ \simeq \
10^{16(2/d -1)} \, \left(\frac{{\rm TeV}}{M_D}\right)^{1+2/d} \, {\rm mm}\, .
\end{equation}
Within length scales of order $R$ gravity is genuinely $4+d$
dimensional and the gravitational potential starts growing as $\sim
1/r^{1+d}$. This rules out $M_D\sim1$~TeV for $d=1$ models, while
leaving a small window for $d=2$ \cite{Long:2003dx}.  A fundamental
Planck mass of $\sim$ TeV opens the intriguing possibility of
recording quantum gravity signals at the next generation of
accelerators, where those energy scales will be actually probed
\cite{QG}.

In this note we investigate whether stronger gravity on sub-mm scales could
affect the thermal history of the early Universe due to the enhanced
gravitational scattering cross-section predicted by these models, even
at energies and temperatures much less than the electroweak
scale. Below the quantum gravity regime, we consider relativistic
scattering processes between species of very different masses $m \ll
M$ in the limit of small scattering angle/low momentum transfer and
estimate the efficiency of such processes for establishing statistical
equilibrium between species.  More specifically, we will consider the
heavy species as relativistic and in thermodynamical equilibrium with
an average energy per particle $\sim T$ and the light species
initially decoupled and ``cold'' i.e. with an energy per particle $\ll
T$.  This is the case of the QCD axions, which acquire an effective
mass $m \simeq 10^{-5} eV$ at temperatures as high as a GeV.

Ultra-light weakly interacting massive particles (WIMPs) such as the
axion are candidates for the cold dark matter (CDM) that is
responsible for around 25\% of the energy density of the Universe
today and play a central role in cosmological structure formation. A
crucial requirement for these ultra-light WIMPs is that they are
non-thermal. In particular the QCD axions \cite{axions} arise from
coherent oscillations about the minimum of its potential of the
spatially homogeneous axion field that begin when the temperature
drops below about 1 GeV at the QCD phase transition. They are supposed
to be effectively non-interacting at this temperature and thus remain
at rest until local overdensities undergo gravitational collapse,
beginning the process of structure formation, from the bottom
up. Today axions are supposed to be present in the virialised dark
halo surrounding visible stars in galaxies, with velocity dispersion
of order $100$~km~s$^{-1}$.

The gravitational cross-section of ultra-light WIMPs, enhanced by the
presence of large extra dimensions, has recently been proposed
\cite{Qin} as a natural mechanism for the CDM ``self-interaction'',
advocated in \cite{Spergel} to resolve the ``cusp problem''
\cite{crisis} in the centre of galaxies (other proposed solutions are
found e.g. in \cite{others}).  In this paper we consider another
effect of the enhanced gravitational scattering. We consider whether
gravitational scattering between ultra-light WIMPs and much heavier
but relativistic species in the early universe (such as electrons or
neutrinos) could establish a statistical equilibrium between different
species at energies well below the fundamental Planck scale. If the
energy transfer is sufficient to make the axions highly relativistic,
then they would no longer be viable candidates for the CDM in models
with large extra dimensions.

\section{Classical gravitational scattering in higher dimensions on a brane}

In the rest frame of the heavy particle, the light particle feels a
central attractive gravitational force
\begin{equation}
 \label{force}
|{\bf F}(r)| \ =\ \frac{M m \bar{\gamma}}{M_D^{d+2} r^{d+2}}, \qquad
 \quad M_D^{-1}  <  r < R\, , 
\end{equation}
where $\bar{v}$ is the speed of the light particle in this frame, and
$\bar{\gamma} \equiv (1 - \bar{v}^2)^{-1/2}$ is the usual relativistic
Lorentz factor. Note that we use this equation as our definition of
$M_D$. Equation~(\ref{force}) is derived from the $4+d$ dimensional
propagator and the lower bound $r > M_D^{-1}$ is the limit of validity
of such a tree-level calculation.  Above the compactification radius,
$r>R$, on the other hand, the usual Newtonian inverse-square law is
restored (or $r>\ell_{AdS}$ in the Randall-Sundrum model \cite{RS}
with an Anti-de Sitter bulk if $R>\ell_{AdS}$).

In the small scattering angle limit the transverse momentum $p_T$
acquired by the light particle can be estimated as
\begin{equation}\label{pt}
p_T \ = \ \frac{2 b}{\bar{v}}\int_0^\infty dx \frac{|{\bf
 F}|}{\sqrt{b^2 + x^2}} \ \simeq 
 \ \frac{\bar{\gamma}}{\bar{v}} \frac{M m}{M_D^{d+2} b^{d+1}},
\end{equation}
where $b$ is the impact parameter.
It will be useful to define the velocity-independent dimensionless parameter 
\begin{equation} \label{epsilon}
\epsilon \ =\ \frac{\bar{v}\, p_T}{\bar{\gamma} m}\  = \
\frac{M}{M_D^{d+2} b^{d+1}} 
\end{equation}
that, in the relativistic limit $\bar{v} \simeq 1$, is just the
transverse momentum in units of the energy of the particle. As long as
we consider impact parameters bigger than the Planck scale cut-off
$M_D^{-1}$, we have $\epsilon < M/M_D \ll1$.

In Appendix~A we calculate the energy transfered to ultra-light WIMPs
due to gravitational scattering with much heavier, but relativistic
particles such as electrons with Lorentz factors $\gamma \gg 1$ in the
cosmological reference frame. We find, from Eq.~(\ref{dEdt}), an
energy transfer rate
\begin{equation}
 \label{transfer}
\frac{dE}{d t}\ \simeq\ \pi \, n_* E \left[\gamma^2 - 1 \right]
 \int_{b_{UV}}^{b_{IR}} db\, b\,     \epsilon^2  
 \, ,
\end{equation}
where $n_*$ is the number density of the heavy particles and
$\epsilon(b)$ is given by Eq.~(\ref{epsilon}). The energy transfer
rate in higher-dimensional gravity [$d>0$ in Eq.~(\ref{epsilon})] is
dominated by scattering events with small impact parameters due to the
steep rise in the gravitational force on small scales $r\ll R$. Thus
the energy transfer is not sensitive to the IR cut-off, $b_{IR}$ (here
given by the compactification scale $\sim R$), but is sensitive to the
UV cut-off, $b_{UV}$.
Note that in 4 dimensional theory ($d=0$) the energy transfer depends
on the cut-off scales only logarithmically $\log (b_{IR}/b_{UV})$.

We must cut-off our perturbative calculation at least at the
fundamental Planck scale, $b_{UV}\sim M_D^{-1}$, which is the limit
down to which we can trust Eq.(\ref{force}).  Note that for elementary
particles with $M < M_D$, the Schwarzchild radius $r_S \sim M_D^{-1}
(M/M_D)^{1/(d+1)}$ of the heavy particle is much smaller than
$M_D^{-1}$. A more serious problem is that our result is sensitive to
trans-Planckian scattering and a reliable calculation requires
knowledge of non-perturbative quantum gravity on smaller scales! 
{\em The classical energy transfer rate due to gravitational
scattering in brane-world models is sensitive to trans-Planckian
physics.}

On the other hand the Compton wavelength of our particles are much
larger than $M_D^{-1}$ so a full calculation requires a quantum
mechanical treatment. We estimate the effect of quantum corrections in
the following section, but in the rest of this section consider the
effect of classical scattering.

We can estimate the classical rate of energy transfer to the
ultra-light particles as 
\begin{equation} \label{rate}
\frac{d E}{dt} \ \simeq \ \pi n_* E [\gamma^2-1] \int_{M_D^{-1}}^R b
\epsilon^2 db \ \simeq \ 
\frac{\pi}{2 d}  \frac{E T^2 n_*}{M_D^4}.
\end{equation}
where we take the heavy species to be relativistic ($v \simeq 1$) and
in thermodynamical equilibrium ($\gamma M\simeq T$). Note that this
result is independent of the rest mass of the heavy particles and
depends only upon the number density of relativistic particles.

Thus we would expect a rapid rise in the energy of ultra-light WIMPS
due to classical gravitational scattering at sufficiently early times
when $\dot{E}/E \gg H$. Note that Eq.~(\ref{rate}) has been derived in
the limit $\bar\gamma m\ll M$, which was used in Appendix A to assume
negligible recoil of the heavy particles during each collision. In
practice the heavy particles are not infinitely heavy, and hence are
not an infinite energy source, and the energy transfer rate must
decrease as statistical equilibrium is approached.


{}From \eqref{rate} the relaxation rate of our light species can be 
expressed as a function of the temperature by taking the number density
$n_* = 1.8 {\cal N}_f T^3/\pi^2$ of fermionic relativistic species,
\begin{equation} 
\frac{\dot{E}}{E} \ = \ 10^{-12}\frac{{\cal N}_f}{\pi d}
\left(\frac{{\rm TeV}}{M_D}\right)^4 
\left(\frac{T}{{\rm GeV}}\right)^5 {\rm GeV}\, ,
\end{equation}
${\cal N}_f$ being the effective number of relativistic fermionic
species.  Gravitational interactions can efficiently establish
equilibrium if, at some epoch in the early Universe, $\dot{E}/E \gg
H$, where the Hubble rate in the radiation dominated era is given by
\begin{equation}
 \label{hubble}
H \simeq 10^{-18}\left(\frac{T}{{\rm GeV}}\right)^2 {\rm GeV}.
\end{equation} 
A more detailed calculation (see Appendix B) shows that the light
particles acquire enough energy so that they are still relativistic at
the beginning of matter domination $T_{\rm eq} \sim 10$ eV, if
$\dot{E}/E > \alpha\, H$, where $\alpha \simeq 30$.  In this case even
if the light particles are initially ``cold'' when produced at $T_A$,
they soon become relativistic due to the gravitational scattering by
heavy relativistic particles in thermal bath. They remain relativistic
until matter-radiation equality if
\begin{equation}
\frac{{\cal N}_f}{\pi d} 
\left(\frac{T_A}{{\rm GeV}}\right)^3 > 10^{-2} \alpha \
\left(\frac{M_D}{10\ {\rm TeV}}\right)^4 \,. 
\end{equation}
Thus, for $M_D \lesssim 10$ TeV, axions produced at $T_A \sim$ GeV
would acquire sufficient energy that, even by the time of matter-radiation
equality, they are still relativistic and would therefore be ruled out
as dark matter candidates.
\footnote{Note that our assumption of elastic scattering of the
ultra-light particles in the rest frame of the heavier particles
required $\bar\gamma m<M$ which corresponds to $E<M^2/T$. For instance
considering gravitational scattering with electrons at $T\sim 1$~GeV
our calculation requires $E/T<10^{-6}$. This is just sufficient to
show that axion with mass $m\sim 10^{-5}$~eV remain relativistic at
$T_{\rm eq}\sim 10$~eV. In practice we don't expect energy transfer to
shut-off completely for $M^2/T<E<T$, but it may become less
efficient.}

The reason why gravitational scattering remains efficient far below the 
fundamental Planck scale is the hierarchy between the fundamental Planck scale 
and the effective four-dimensional Planck scale, $M_P\sim10^{19}$~GeV, which 
determines (in practice, suppresses) the Hubble expansion rate at these 
energies. For models where the fundamental Planck scale is as low as a TeV, 
gravitational scattering of ultra-light WIMPS remains efficient down to 
temperatures as low as $10$MeV. This is analogous to the neutrinos remain 
coupled to baryons down to energies of order $1$~MeV even though the 
electroweak scale is of order $100$~GeV.

\section{Quantum effects}

In calculating the energy transfer in the previous section
we considered the particles as point-like and the scattering as a
completely classical processes. By taking $b_{UV}$ in \eqref{transfer}
as $M_D^{-1}$ we have implicitly assumed that the two particles can
come arbitrarily closed to each other, neglecting their finite Compton
wavelengths. In the mass frame of the heavy particle the Compton
wavelength of the light particle is $(\bar{\gamma} m)^{-1}$. By taking
$\bar{\gamma}$ as given by \eqref{gamma} and by averaging over the angle
$\theta$ we can estimate the Compton wavelength of the light particle
as $\sim M/ (E T)$. By taking this rather than the Planck scale
as the small scale cut-off $b_{UV}$ in
\eqref{transfer} we obtain the energy transfer
\begin{equation} \label{rate2}
\frac{d E}{dt} \ \simeq \ \frac{E^{2 d+1} T^{2d +5}}{M^{2d} M_D^{2d+4}} ,
\end{equation}
here we assume $R>(\bar{\gamma} m)^{-1}$ so that we still probe the 
higher-dimensional gravity.
The relaxation rate $\dot{E}/E$ for axions is now much smaller
that the Hubble rate $H$ for $M_D\sim $TeV:
\begin{equation}
\frac{\dot{E}/E}{H} \ \simeq \ \frac{M_P E^{2 d} T^{2d +3}}{ M^{2d}
  M_D^{2d+4}} \ \simeq \  
10^{7 - 6 d} \left(\frac{T_A}{GeV}\right)^{2d +3}
  \left(\frac{TeV}{M_D}\right)^{2d +3}  
\left(\frac{m}{M}\right)^{2d} .
\end{equation}
The last equality holds at $T = T_A$ when the energy of the axions is
$E \simeq m$. 

Rather than introducing an abrupt cut-off at the Compton wavelength, we
can try to estimate the suppression due to the interference of the
wavefunction for impact parameters smaller than the Compton wavelength,
$l_C$, of the particle.  In the case of a uniform potential well of
radius $a$ the quantum mechanical cross section is suppressed, with
respect to the classical one, by a factor $(a/l_C)^6$ \cite{book},
i.e.  $\sigma_{\rm quantum} = \sigma_{\rm classical} (a/l_C)^6$. The
cross section has a direct interpretation in terms of ``number of
particles scattered''. Therefore, in the integral in \eqref{rate},
when considering impact parameters smaller than the Compton
wavelength, $b<l_C$, we should include the corresponding suppression
factor:
\begin{equation}
\frac{d E}{dt} \ \simeq \ \pi n_* E [\gamma^2-1] \int_{M_D^{-1}}
\left(\frac{b T E}{M}\right)^6 
b\, \epsilon^2 db \ \simeq \ \frac{E^7 T^{11}}{M^6 M_D^{2d + 4}}
\int_{M_D^{-1}}  
b^{-2 d + 5} d b. 
\end{equation}
Note that, as recently emphasized also in \cite{Chatillon}, for $d
\geqslant 3$ the steepness of the gravitational potential
wins over the quantum suppression effect, 
and the process is still sensitive to trans-Planckian
physics. For $d<3$, on the other hand, most of the contribution comes
from impact parameters larger than the Compton wavelength. Note also
that for $d \geqslant 3$ the energy rate is given by \eqref{rate2}
with $d = 3$.

\section{Conclusion}

In this letter, we studied the impact of an enhanced gravitational
scattering cross-section in the early universe in the context of the
brane-world models with a low fundamental Planck scale. In the
presence of a general interaction of some given elastic scattering
differential cross section, we calculated the relativistic energy
transfer rate for two species with a large ratio between their masses.
Due to the enhanced gravitational scattering cross section,
ultra-light WIMPs could reach a statistical equilibrium with heavier
relativistic particles in the thermal bath even well below the
fundamental Planck scale.  For example, the QCD axions produced at
$\sim$ 1 GeV could soon become relativistic and they remain
relativistic until matter-radiation equality due to classical
gravitational scattering.
Axions are usually supposed to be non-interacting because their
interactions are suppressed due to the existence of a much larger mass
scale $f_a\sim 10^{12}$~GeV. But in models with large extra dimensions
their gravitational interactions can only be suppressed by the
fundamental Planck scale, $M_D\ll f_a$.

However the classical energy transfer is dominated by scattering
events with impact parameters much less than the Compton wavelength of
the light particles.
If we cut-off our classical calculation at the Compton wavelength the
effect disappears completely . If instead we attempt to model the
quantum suppression of the calculation below the Compton wavelength we
find that for $d>3$ the result remains sensitive to the ultra-violet
cut-off of the theory at the fundamental Planck scale
\cite{Chatillon}. This is due to the steep rise in the gravitational
force on small scales in higher dimensional spacetime.  Thus, in a
higher dimensional theory, the energy transfer due to gravitational
scattering is sensitive to trans-Planckian physics. 

It is interesting to consider whether other phenomena could be
sensitive to enhanced gravitational scattering. We commonly neglect
the effect of gravitational scattering as we expect it to be weak
compared to all other interactions. But in models with large extra
dimensions gravity can be much stronger on small scales than we
naively imagine. And in contrast to four-dimensional gravity, the
scattering is sensitive to trans-Planckian physics. Even the absence
of detectable gravitational interactions could place constraints upon
gravitational scattering on sub-Planck scales.

\section*{Acknowledgments} It is a pleasure to thank Paolo Nason, 
Gabriele Veneziano and Alberto Zaffaroni for useful discussions and
correspondence.

\appendix

\section{Energy transfer rate by classical relativistic scattering}

We want to study the energy transfer between two particle species,
with masses $m_1 \gg m_2$, in terms of classical relativistic
scattering. We assume that in the cosmological reference frame
(e.g. the rest frame picked out by the cosmic microwave background)
these species have speeds $v_1$ and $v_2$, and that in this cosmic
rest frame the velocity distribution of both species are isotropic.

A given scattering event is characterized by the angle $\theta$
between the initial 3-velocities ${\bf v}_1$ and ${\bf v}_2$ of the
two particles considered.  Without loss of generality we can write the
initial four-momenta in the cosmic rest frame as $m_1 \gamma_1 (1, \
v_1,\ 0,\ 0)$ and $m_2 \gamma_2 (1, \ v_2 \cos \theta, \ v_2 \sin
\theta,\ 0)$ respectively, where $\gamma_1$ and $\gamma_2$ are the
usual relativistic Lorentz factors.

In the particle-1 rest frame (RF1), the four-momentum of particle 2 is
\begin{equation}
\bar{p}_{\rm ini} \equiv
 (\bar{E}, \ \bar{p}_x, \ \bar{p}_y, \ \bar{p}_z)\ 
\equiv \ m_2 \gamma_2 \left(\gamma_1 [1-v_1 v_2\cos\theta],\ \gamma_1 
[-v_1 +v_2 \cos \theta], 
\ v_2 \sin \theta,\ 0\right), 
\end{equation}
from which the Lorentz factor $\bar{\gamma}$ and relative velocity $\bar{v}$ in the RF1 frame are derived as 
\begin{equation}
 \label{gamma}
\bar{\gamma} (\theta) \ \equiv \ \frac{1}{\sqrt{1-\bar{v}(\theta)^2}}\ = \ \gamma_1 \gamma_2 
(1-v_1 v_2 \cos\theta).
\end{equation}

In RF1 we can characterise the geometry of each scattering event by
the two parameters $b$ and $\Phi$. We assume that we know the
scattering deflection angle $\Theta(b,\bar{v})$ in RF1 as a function
of the impact parameter $b$ and velocity $\bar{v}$.  (In the case of
interest in this paper $\Theta \simeq \epsilon/\bar{v}^2 \ll 1$ and
$\epsilon$ is given in equation \eqref{epsilon} as a function of $b$).
After scattering, the light particle can acquire an orthogonal
velocity component.  We define an angle $\Phi$ as the angle between
the plane $x-y$ and the plane containing 3-velocity of the light
particle after scattering as well as particle 1.  Only for $\Phi = 0$
or $\Phi = \pi$, the particle is scattered on the $x-y$ plane and in
general its velocity acquires an orthogonal component proportional to
$\sin\Phi$.
The final four-momentum after an elastic collision in RF1 is thus given by 
\begin{equation}
\bar{p}_{\rm fin} = \left(\bar{E},\ 
\bar{p}_x \, \cos \Theta + \bar{p}_y\, \sin \Theta \cos\Phi, \  
\bar{p}_y\, \cos \Theta - \bar{p}_x\,  \sin \Theta \cos\Phi,\ 
\sqrt{{\bar{p}_x}^2 +{\bar{p}_y}^2}\, \sin \Theta \sin\Phi 
\right)
\end{equation}
Because we are working in the limit $m_2/m_1\to0$ we require that the
final speed of particle 2 is the same as its initial speed in RF1 and
only its direction is changed. Total momentum is conserved due to the
recoil of particle 1, but the kinetic energy transfered in RF1 is
negligible as $m_2/m_1\to0$.

For each particle of type 1 the differential rate of such an event is
\begin{equation} \label{drate}
d \bar{\Gamma}_1(b,\theta,\Phi) \ = \ b \, \bar{v} (\theta) \, d \bar{n}_2(\theta)\, db\, d\Phi\, ,
\end{equation}
where $d \bar{n}_2$ is the RF1-number density of those type 2
particles whose direction in the CMB reference frame is comprised
between $\theta$ and $\theta + d\theta$.

By Lorentz-boosting back to the cosmic rest frame we obtain the total
energy of the light particle after such an event:
\begin{equation}
E(b, \theta, \Phi)\ =\ m_2  \gamma_1 \left[\bar{\gamma} + \left(\frac{\gamma_2}{\gamma_1} -\bar{\gamma}\right)\cos\Theta   + 
v_1 v_2 \gamma_2 \sin\Theta \cos \Phi \sin \theta  \right] \,.
\end{equation}
Thus the energy acquired after the scattering, $\Delta E \equiv E - m
\gamma_2$, in the limit of small scattering angle, $\Theta \simeq
\epsilon/\bar{v}^2$, is
\begin{equation} \label{deltae}
\Delta E(b, \theta, \Phi)\ =\ m_2  \gamma_1 \left[\frac{\epsilon
    (b)^2}{2 \bar{v}(\theta)^4} \left(\frac{\gamma_2}{\gamma_1}
  -\bar{\gamma}\right) + \frac{\epsilon (b)}{ \bar{v}(\theta)^2} 
v_1 v_2 \gamma_2 \cos \Phi \sin \theta  \right]\, .
\end{equation}

The differential scattering rate in the cosmic rest frame is also
easily worked out from \eqref{drate},
\begin{equation}
d \Gamma_2(\theta, b, \Phi) \ = \ b \, \bar{v} (\theta) \, d
n_1(\theta)\, db\, d\Phi\, . 
\end{equation}
By the assumption of isotropy in the cosmic rest frame
\begin{equation}
d n_1(\theta) \ = \ \frac{n_1}{2} \sin \theta \, d\theta ,
\end{equation}
where $n_1$ is just the number density of species 1.

Finally, the energy transfer rate in the cosmic frame
\begin{equation}
\frac{d E}{d t}\ =\ \int \Delta E d \Gamma
\end{equation}
can be calculated. By integrating over $\Phi$, from $0$ to $2 \pi$,
the $\Phi$-dependent term in \eqref{deltae} averages to zero. The
integration over $\theta$, from $0$ to $\pi$ can be worked out in the
ultra-relativistic limit $\gamma_1 \gg 1$. We are left with
\begin{equation}
 \label{dEdt}
\frac{d E}{d t}\ =\ \pi \, n_1 \gamma_2\, m_2 \left[\gamma_1^2 - 1 + {\cal O}(\gamma_1^{-2}) \right] \int_{b_{UV}}^{b_{IR}} db\, b\, \epsilon^2 
 \, .
\end{equation}
In the case of interest in this paper we must impose a UV cut-off $b_{UV}\sim M_D^{-1}$ below which the classical scattering amplitude will not be valid, and at long wavelengths there is a cut-off $b_{IR}\sim R$ beyond which we recover Newtonian gravity.

\section{Energy evolution of the light species}

We now want to follow the Energy evolution of the axions down from the
temperature $T_A\sim$ GeV where they are given a mass $m \simeq
10^{-5}$ eV and they are basically at rest in the reference frame of
the cosmological observers.  The axions' energy is controlled by two
main processes: the gravitational interaction with some relativistic
species (electrons, positrons or neutrinos) described in the text and
the energy redshift due to the expansion of the Universe:
\begin{equation}
d E\ = \ \left(\frac{{\cal N}_f M_P}{5 \pi d M_D^4} E T^2 - \frac{E}{T}
+\frac{m^2}{ET}\right) (-dT)
\end{equation}
The first term in the parenthesis comes directly from eq. \eqref{rate}, where
temperature has been used instead of time 
($t \simeq 10^{-1} M_P/T^2$ during radiation domination) as independent variable
and $n_* = 1.8\, T^3/\pi^2$ has been also used. The second and third terms 
represent the energy loss by redshift. In the range of interest between
$T_A$ and the beginning of matter domination at $T\simeq 10$ eV the third term 
is irrelevant. By posing $E(T_A) = m$ one finds the solution
\begin{equation}
\frac{E(T)}{m}\ =\ \frac{T}{T_A}\ e^{-{\cal C}(T^3 -T_A^3)},
\end{equation}
where 
\begin{equation}
{\cal C}\ \equiv \ \frac{{\cal N}_f M_P}{15 \pi d M_D^4}\ \simeq \
10^6 \frac{2 {\cal N}_f}{3 \pi d} \left(\frac{{\rm TeV}}{M_D}\right)^4 {\rm GeV}^{-3} .
\end{equation}
The ``heating up'' process of the axions is very efficient at the beginning 
and $E(T)$ reaches a maximum at a temperature which is a fraction of $T_A$. From 
then on, axions cool down by the effect of the redshift; in this regime 
$E$ is basically linear in $T$:
\begin{equation}
\frac{E(T)}{m}\ \simeq \ \frac{T}{T_A}\ e^{{\cal C}T_A^3}\, , \qquad T \ll T_A\, .
\end{equation}
Requiring that the axion be non relativistic at $T_{\rm eq} = 10$ eV implies requiring 
the above number to be of order one at that temperature. We obtain 
${\cal C} T_A^3 > 20$, i.e.
\begin{equation}
\frac{{\cal N}_f}{\pi d} 
\left(\frac{T_A}{{\rm GeV}}\right)^{3}\, > 
3 \times 10^{-5} \left(\frac{M_D}{\rm TeV} \right)^4.
\end{equation}


\end{document}